\documentclass{article}
\usepackage{lnfprep}
\usepackage{graphicx}
\usepackage{subfigure}
\usepackage{amsmath}
\usepackage{amssymb}

\def\beq{\begin{equation}}   \def\eeq{\end{equation}}
\def\bea{\begin{eqnarray}}   \def\eea{\end{eqnarray}}

\newcommand{\gsim}{\lower.7ex\hbox{$ \;\stackrel{\textstyle>}{\sim}\;$}}
\newcommand{\lsim}{\lower.7ex\hbox{$ \;\stackrel{\textstyle<}{\sim}\;$}}
\def\c2{CLEO~II.V}

\def\d0d0{ D^0\bar{D}^0 }
\def\p0p0{ P^0\bar{P}^0 }

\def\qp2{ \Bigl| \frac{q}{p} \Bigr|^2 }
\def\pq2{ \Bigl| \frac{p}{q} \Bigr|^2 }

\def\ps2s{  \psi(2S) }
\def\q2{ $q^2$ }
\def\cm2s1{ $\,{\rm cm}^{-2} {\rm s}^{-1}$}
\def\d0{D_2^{*0}}
\def\d+{D_2^{*+}}

\newcommand{\Header}{
  \begin{tabular}{rl}
  \hspace{-.4cm}
      &
    \renewcommand{\arraystretch}{0.5}
    \renewcommand{\arraystretch}{1}
  \end{tabular}
  \vskip 1cm
  \begin{flushright}
  \renewcommand{\arraystretch}{0.5}
    \begin{tabular}{r}
      {\underline{LNF-04/25 (P)}}\\    % insert here the preprint number
      {\small 15 novembre 2004} \\      % insert here the preprint Date
    \end{tabular}
  \end{flushright}
  \renewcommand{\arraystretch}{1}
  \vskip 1 cm
  }
\begin{document}
\begin{titlepage}
\title{
  \Header
  {\LARGE  \textsc{\textmd {An  Unconventional Approach For A\\
  Straw Tube-Microstrip Detector}}
  }
}
\author{E.Basile(*), F.Bellucci (***), L.
Benussi, M. Bertani, S. Bianco, M.A. Caponero (**),  \\
D. Colonna (*), F. Di Falco (*), F.L. Fabbri, F. Felli (*), M.
Giardoni, A. La Monaca, \\
G. Mensitieri (***), B. Ortenzi, M. Pallotta, A. Paolozzi (*), L.
Passamonti, D.Pierluigi, \\
C. Pucci (*), A. Russo, G. Saviano (*), S. Tomassini.\\
{\em Laboratori Nazionali di Frascati dell'INFN }}
\maketitle
\baselineskip=1pt

\begin{abstract}
\indent \indent We report on a novel concept of silicon microstrips
and straw tubes detector, where integration is accomplished by a
straw module with straws not subjected to mechanical tension in a
Rohacell $^{ \circledR}$ lattice and carbon fiber reinforced plastic
shell. Results on mechanical and test beam performances are reported
on as well.
\end{abstract}

\vspace*{\stretch{2}}
\begin{flushleft}
% insert here the PACS number
  \vskip 2cm
{ PACS.: wire chambers, straw tubes, HEP detectors, silicon
detectors, silicon microstrips, beauty quark, CP violation.}
\end{flushleft}
\begin{center}
\emph{Submitted to Transactions on Nuclear Science}
\end{center}

\vskip 1cm
\begin{flushleft}
\begin{tabular}{l l}
  \hline
  $ ^{*\,\,\,\,\,\,}$ & \footnotesize{Permanent address: ``La Sapienza" University - Rome.} \\
  $ ^{**\,\,\,}$& \footnotesize{Permanent address: ENEA Frascati.} \\
  $ ^{***}$ & \footnotesize{Permanent address: ``Federico II" University - Naples.} \\
\end{tabular}
\end{flushleft}
\end{titlepage}
\pagestyle{plain}
\setcounter{page}2
\baselineskip=17pt

\section{  \textsc{Concept}}
Modern physics detectors are  based on tracking subcomponents, such
as silicon pixels and strips, straw tubes and drift chambers, which
require high space resolution, large geometrical acceptance and
extremely large-scale integration. Detectors are often requested
demanding requirements of hermeticity and \linebreak[4] compactness
that must satisfy the minimization of materials. We have developed
an integration solution that accommodates straw tubes
and silicon strips in a common structure.\\
Our novel design utilizes straw tubes mechanically non-tensioned and
embedded in a Rohacell $^{ \circledR}$ lattice.
\newline
\section{  \textsc{BTeV Detector}}

Experiment BTeV[1] at the Fermilab proton-antiproton collider
Tevatron produces and studies the \linebreak[4] elementary particles
composed of the heavy quark beauty, in order to investigate the
phenomenon called CP violation, and understand if the Standard Model
of particles and interactions is sufficient to describe the world we
live in. BTeV is composed of tracking detectors  (pixel, strips,
straws) for detection of charged particles, RICH Cerenkov detector
for identification of pions, kaons and protons, crystal EM
calorimeter for detection of neutral particles (photons and
$\pi^0$),
 and muon detector.
 \newline
\section{  \textsc{M0 Concept}}
M0 is a special straw module which houses straw tubes and supports
silicon microstrip detectors planes.  M0 is made of straw tubes
embedded in a Rohacell $^{ \circledR}$ foam, inside a Carbon Fiber
Reinforced Plastic (CFRP) shell. CFRP is chosen to allow the
fabrication of a rigid mechanical structure with high transparency
to incoming particles. CFRP is also used for M1 modules,
conventional straw tubes sub detectors that act as struts sustaining
the mechanical tension of remaining straw modules. Six
straw-microstrips stations are deployed in BTeV, each station made
of three views, each view made of two half-views. Straw lengths vary
from $54 cm$ in the first station to $231 cm$ in the sixth station.
\newline
\section{  \textsc{FEA Validation}}
A Finite Element Analysis (FEA) of these structures allows us to
estimate the displacements of the \linebreak[4] M0 and M1 modules
under the loads of the micro-strips and straws tubes. Time stability
and maximum displacements of the order of $10 \mu m$ are requested,
in order not to spoil the space resolution of microstrip detectors.
The FEA analysis has been carried on the M0 of the sixth station,
the longest straw length.\linebreak[4] A straw load of $1.4N$ in
each corner of M0 has been simulated to reproduce the mechanical
tension of wires. A load of $12.3N$ and a momentum of torsion of
$2Nm$ have been applied to simulate the weight of the  micro-strip.
The used material properties are reported in the table below. FEA
shows a maximum displacement of $15 \mu m$ \, ($4 \mu m$ in the
axial direction, $2 \mu m$ x direction, $11 \mu m$ y direction),
close to the \linebreak[4] required specification. We have used
shell elements for the simulation of the carbon fiber reinforced
\linebreak[4] polyester structure and bricks for the Rohacell $^{
\circledR}$ simulation. Beam elements were used for introducing glue
between the CRFP modulus and the cylindrical plate where microstrips
were placed.
\newline
\begin{table}[!h]
\begin{center}
\begin{tabular}{|c||c|c|c|}
\hline

 & M\textsc{0 and M1 Structure} & \textsc{Micro Strip Cylinder}  & \textsc{Rohacell Foam} \\
%\hline
\hline
\hline
\textsc{Thickness} & 0.07 \textsc{each ply with} & 0.07 \textsc{each ply} (0°/90°/0°) & \\
%\hline
\textsc{[mm]} & \textsc{fibres disposition} & \textsc{with a rohacell}& 2 \\
%\hline
  & \textsc{of} 0°/90°/0° & \textsc{foam of 5cm}&  \\
\hline
$E_{11}[GPa]$ & 260 & 590 & 0.019 \\
\hline
$E_{22}[GPa]$ & 10 & 10 & - \\
\hline
$E_{12}[GPa]$ & 7.2 & 7.2 & - \\
\hline
$\nu_{12}$ & 0.3 & 0.3 & 0.3 \\
\hline
\end{tabular}
\caption{Geometrical arrangement (thicknesses), Young modules
($E_{11}$, $E_{12}$, $E_{22}$), and Poisson coefficient $(\nu_{12})$
for a FEM simulation of M0 and M1.}
\end{center}
\end{table}
\section{  \textsc{Tomography and FBG}}
A check of the eccentricity of the straws and of their positions in
the grooves can be done with tomography method. The tomography uses
X-ray and can reconstruct sections of the scanned region. Due to the
short X-ray wavelength of about $0.1nm$, the technique determines
the amount of inner surfaces and interfaces of micrometer
dimensions. Computed images are reconstructed from a large number of
measurements of X-ray transmission. The result images are
bidimensional, but a 3-D image is allowed using a digital
\linebreak[4] reconstruction. Results show how a precision of about
20 $\mu m$ can be reached on the measurement of straws radii. The
maximum variation from circularity allowed is 100 $\mu m$. The BTeV
detectors utilize Fiber Bragg Grating (FBG) sensors to monitor
online the positions of the straws and microstrip. The optical fiber
\linebreak[4] is used for monitoring displacements and strains in
mechanical structures such as the presented straw \linebreak[4]
tubes-microstrip support. A wavelength selective light diffraction
along the FBG sensor is placed in the fiber, and it permits an
on-time monitoring of the support. According to these proprieties,
an FBG sensor is going to be placed in the M0 and M1 structure
between the Rohacell $^{ \circledR}$ foam and the CFRP shell.
\newline
\section{  \textsc{Prototype}}
MOX prototypes have been fabricated in order to study the
construction procedures, mechanical properties,\linebreak[4]
material characterization, and physical behaviour for detection of
particles, in test beam set-ups.  The most \linebreak[4] demanding
design requirement is the assembly of straws in a close pack, with
no mechanical tension \linebreak[4] applied. Several gluing
techniques have been examined and tested to determine the optimal
technique. Straw tubes are glued together in three layers, and the
upper and lower layer are glued to the Rohacell $^{ \circledR}$
foam. Glues with different viscosity, and several gluing techniques,
have been used. Glues tested range from cianoacrylate (Loctite 401)
to epoxy (Eccobond series). Gluing techniques ranged from brush, to
injection, to spray gluing. The most promising results have been
obtained by using Eccobond 45W and \linebreak[4] catalyst mixture
(1:1 by weight), diluted with dimethylcheton solvent. For each $40g$
of glue-catalyst \linebreak[4] mixture, $40cm^{3}$ of solvent was
used. The assembly process proceeds as follows. Stainless steel rods
are inserted in each straw tube. A straw layer is formed by locating
48 straws on machined grooved plate.  The glue-solvent mixture
described is sprayed, with $2bar$ air pressure, and $20cm$ distance
between spray\linebreak[4] gun and straw layer. After curing at room
temperature, straw layers are sprayed again and layers are
\linebreak[4] superimposed. After additional curing, stainless steel
rods are removed from straws. Conductive contact is accomplished via
spraying of Eccobond 57C.
\newline
\section{\textsc{Cosmic Ray And Test Beam Results}}
Preliminary results with cosmic rays show very clean pulses in gas
mixtures of interest for BTeV \linebreak[4] (Ar-CO$_{2}$ 80/20), as
shown in Fig.5. Prototypes have been exposed to beam particles in
the Frascati Test \linebreak[4] Beam Facility [7]. Preliminary
results show the expected response of prototype to minimum ionizing
\linebreak[4] particles. The distribution of drift times of gas ions
to the straw wire (Fig.6) over the straw $2mm$ \linebreak[4] radius
is compatible with the drift velocity in Ar-CO$_{2}$ (80/20) gas
mixture used.
\newline
\section{  \textsc{Conclusions}}
We have developed a novel concept for integration of straw tubes
tracking detectors and silicon microstrip detectors, for use in HEP
experiments at hadron colliders. In our design, silicon microstrips
are integrated to a straw tube special module MOX via a CFRP
mechanical structure. Detailed Finite Element Analysis shows that
deformations affect negligibly the tracking performances of the
system. A complete system based on Fiber Bragg Grating sensors ---
acting as optical strain gauges --- monitors the position of each
sub detector with a micron-resolution. The special straw tube module
MOX is realized via straws embedded in a Rohacell $^{ \circledR}$
lattice with no need of mechanical tension. Preliminary results show
that the MOX can provide the 100 $\mu m$ resolution needed by the
BteV tracking detector requirements.
\newline
\section*{  \textsc{Acknowledgements}}
\indent We gratefully thank F.Baruffaldi and his team  at Laboratori
Ortopedici Rizzoli (Bologna, Italy) for help and advice on
tomographic imaging of MOX prototype. We also thank G.Mazzitelli
(LNF INFN, Italy) and all the DA$\Phi$NE team for smooth running of
the Beam Test Facility.
\newpage

\newpage
\begin{figure}[!htbp]
\begin{center}
  \includegraphics[width=10cm]{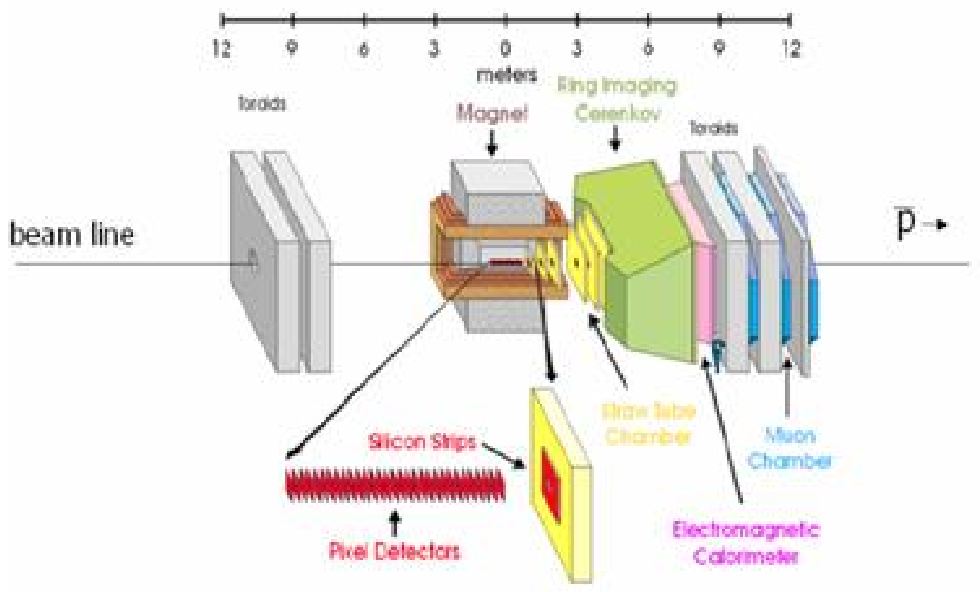}\\
  \caption{BTeV detector layout}
  \end{center}
\end{figure}

\

\begin{figure}[!htbp]
\begin{center}
  \includegraphics[width=15cm]{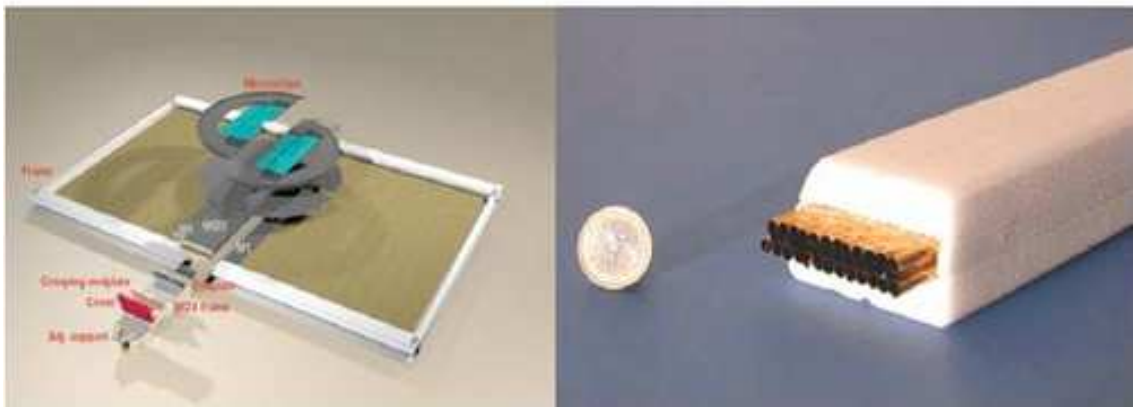}\\
  \caption{BTeV microstrip and straws tubes integration (left);
  MOX prototype with straw tubes embedded in  Rohacell $^{
\circledR.}$}
  \end{center}
\end{figure}

\begin{figure}
\begin{center}
  \includegraphics[width=15cm]{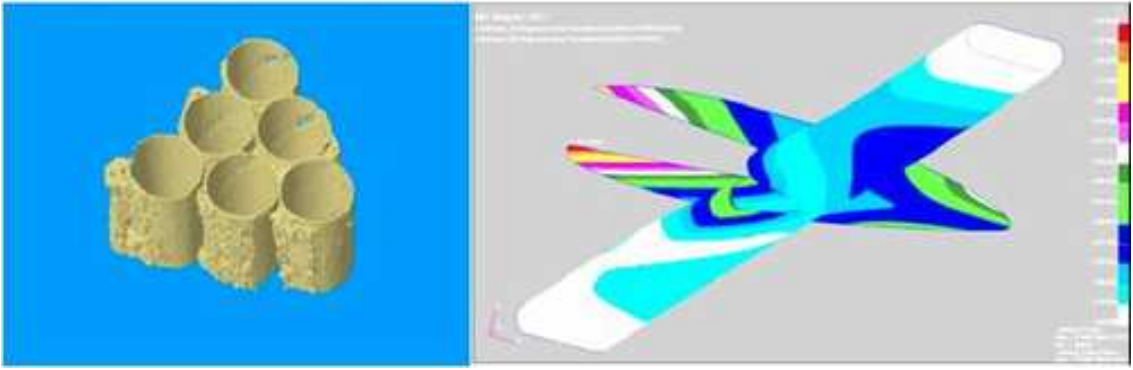}\\
  \caption{Tomography 3D reconstruction of MOX prototype (left);
  FEA results on MOX CFRP shell(right).}
  \end{center}
\end{figure}

\begin{figure}
\begin{center}
  \includegraphics[width=10cm]{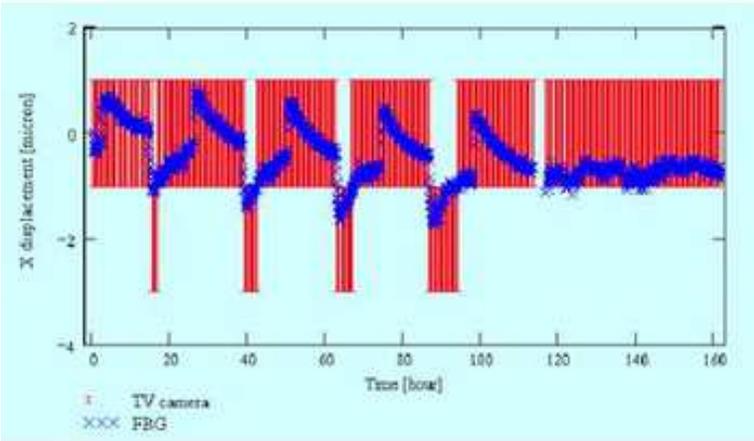}\\
  \caption{FBG long-term monitoring stability results.
  FBG output (crosses)  is validated by TV camera (bars).}
  \end{center}
\end{figure}

\begin{figure}
\begin{center}
  \includegraphics[width=5cm]{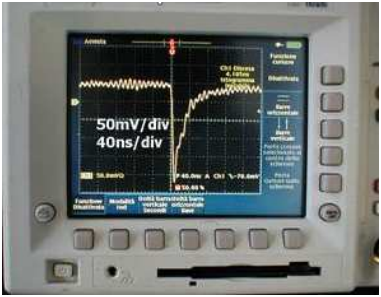}\\
  \caption{Cosmic rays signals in M0X prototype.}
  \end{center}
\end{figure}

\begin{figure}
\begin{center}
  \includegraphics[width=15cm]{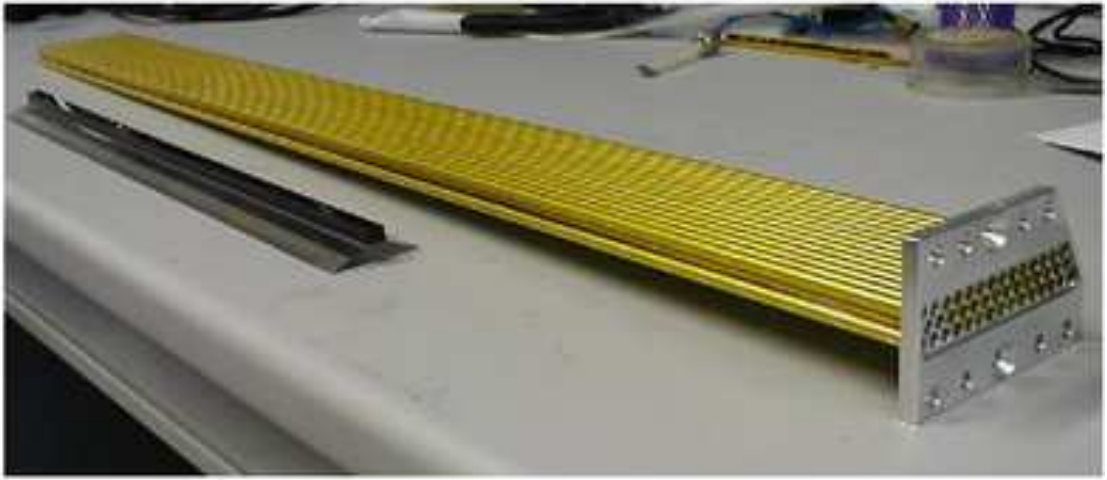}\\
  \caption{MOX module prototype. Straw tubes are glued together
  and positioned between end-plates (one shown) without mechanical tension.}
  \end{center}
\end{figure}

\begin{figure}
\begin{center}
  \includegraphics[width=5cm]{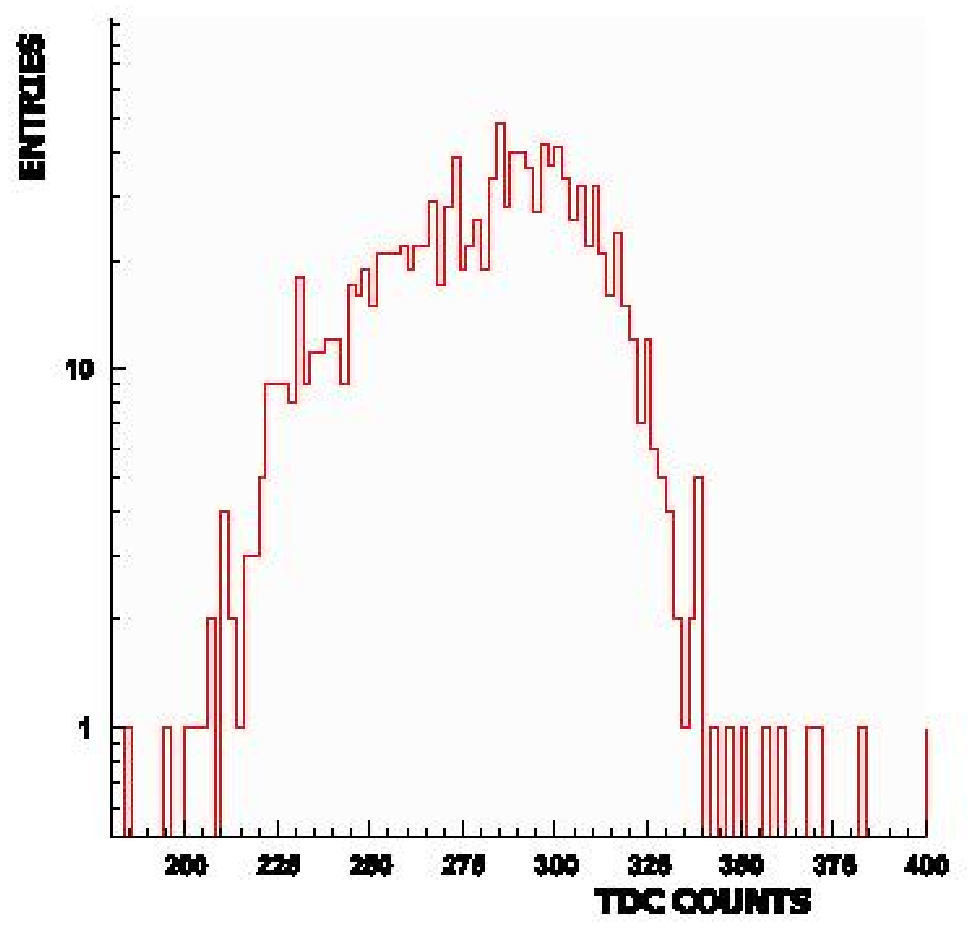}\\
  \caption{Arrival times for beam particles to MOX wire. Times are expressed
in Time-To-Digit-Converted channels (300ps/count).}
  \end{center}
\end{figure}


\begin{thebibliography}{0}

\bibitem
.www-btev.fnal.com

\bibitem
.E.Basile F. Bellucci M.Bertani S.Bianco M.A.Caponero F. Di Falco
F.L.Fabbri F.Felli M.Giardoni A. La Monaca G.Mensitieri B.Ortenzi
M.Pallotta A.Paolozzi L.Passamonti D.Pierluigi A.Russo S.Tomassini,
"Study of Tensile Response of Kapton, and Mylar Strips to Ar and CO2
Mixtures for the BTeV Straw Tube Detector", presented by F. Di Falco
at 10th Vienna Conference On Instrumentation 16-21 Feb 2004, Vienna,
Austria, LNF - 04 / 5(P).
\bibitem
.L. Benussi et al., "Results of Long-Term Position Monitoring by
Means of Fiber Bragg Grating Sensors for the BTeV Detector", LNF -
03 / 15(IR)

\bibitem
.S. Berardis M. Bertani S.Bianco M.A.Caponero F. Di Falco F.L.Fabbri
F.Felli M.Giardoni A. La Monaca B.Ortenzi E. Pace M.Pallotta
A.Paolozzi S.Tomassini, "Fiber optic sensors for space missions"
2003 IEEE Aerospace Conference Proceeding, Big Shy Montana, March
8-15, 2003, pp. 1661-1668

\bibitem
.E. Basile, "Scelta dei materiali ed analisi strutturale per
supporti di rivelatori di particelle dell'esperimento BTeV a
Fermilab (U.S.A.)", degree thesis, University "La Sapienza", Rome,
2003.

\bibitem
.C. Pucci, "Analisi strutturale del supporto per microstrip straw
tubes per l'esperimento di fisica delle particelle BTeV", degree
thesis, University "La Sapienza", Rome, 2004.

\bibitem
.G.Mazzitelli, A.Ghigo, F.Sannibale, P.Valente,       G.Vignola,
Nucl. Instr. Methods A515 (2003) 524-542.



\end{thebibliography}
\end{document}